**Charge Transfer from Perovskite Quantum Dots to Multifunctional Ligands with Tethered Molecular Species**

*Mariam Kurashvili,[1▲] Lena S. Stickel,[1▲] Jordi Llusar,[2] Christian Wilhelm,[3] Fabian Felixberger,[1] Ivana Ivanović-Burmazović,[3] Ivan Infante,[2,4]* Jochen Feldmann,[1]* Quinten A. Akkerman[1]**

[1] Chair for Photonics and Optoelectronics, Nano-Institute Munich, Department of Physics, Ludwig-Maximilians-Universität (LMU), Königinstraße 10, 80539 Munich, Germany

[2] BCMaterials, Basque Center for Materials, Applications, and Nanostructures, UPV/EHU Science Park, Leioa 48940, Spain

[3] Department of Chemistry, Ludwig-Maximilians Universität (LMU), Butenandstraße 5−13, Haus D, 81377 München, Germany

[4] Ikerbasque Basque Foundation for Science, Bilbao 48009, Spain

▲ equal contribution

Correspondence: ivan.infante@bcmaterials.net, feldmann@lmu.de, q.akkerman@lmu.de

**Abstract**

Perovskite quantum dots (pQDs) are promising materials for optoelectronic and photocatalytic applications due to their unique optical properties. To enhance charge carrier extraction or injection donor/acceptor molecules can be tethered to the pQD. These molecules must strongly bind to the ionic surfaces of pQDs without compromising colloidal stability. These we achieve by using multifunctional ligands containing a quaternary ammonium binding group for strong pQDs surface attachment, a long tail group for colloidal stability, and a functional group near the pQD surface. Such pQDs with ferrocene-functionalized ligands show fast photoexcited hole transfer with near-unity efficiency. Density functional theory calculations reveal how ferrocene's molecular structure reorganizes following hole transfer, affecting its charge separation efficiency. This approach can also be extended to in photoexcited electron and energy transfer processes with pQDs. Therefore, this strategy offers a blueprint for creating efficient QD-molecular hybrids for applications like photocatalysis.

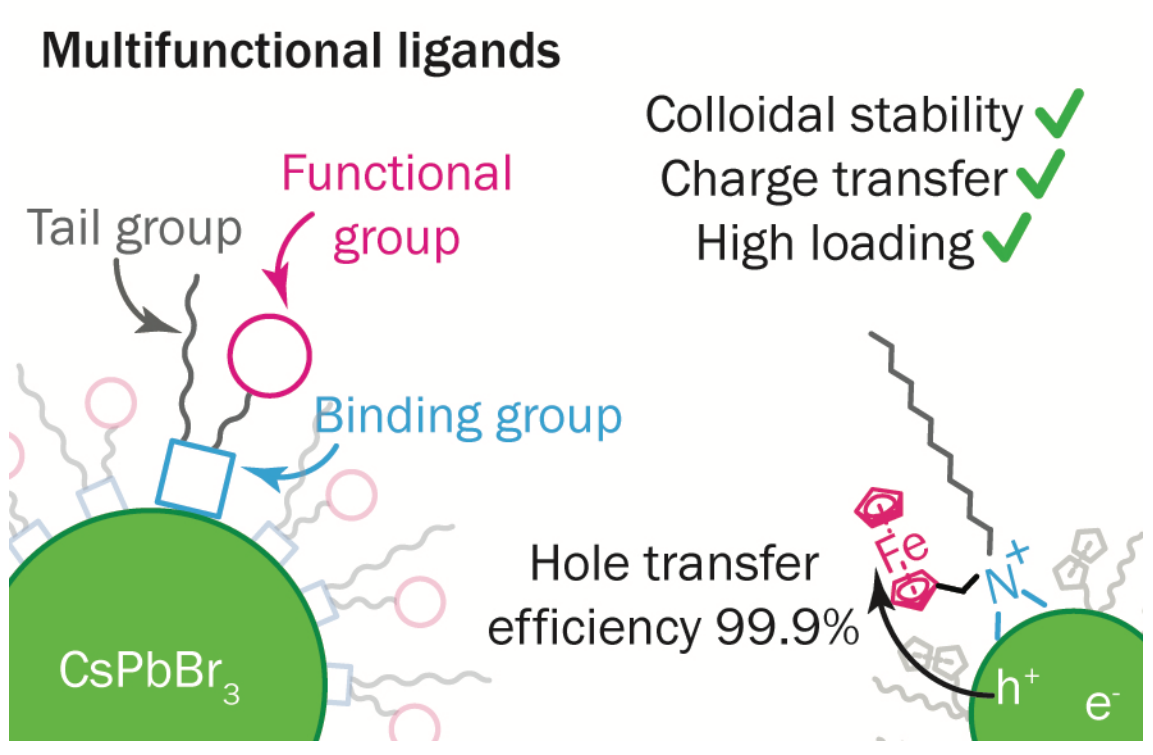

Collodial quantum dots (QDs) are promising materials for applications in optoelectronics,[1, 2] photovoltaics,[3] quantum communication,[4] and photocatalysis.[5] For many of these applications, it is crucial to efficiently inject or extract carriers from the QDs, which remains a challenge due to their organic ligand shell. To facilitate the carrier extraction and injection, molecular species that act as charge or energy donors/acceptors for the QDs, can be added to the ligand shell, enabling charge carrier separation or energy funneling. Such hybrid QD-molecular systems can exhibit efficient charge transfer (CT) used for photocatalysis or solar cells,[6-10] or energy transfer (ET), important for LEDs and sensing applications.[11-13]

To design efficient QD-molecule hybrids, the separation between the QDs and acceptor/donor molecules needs to be minimized, as the interaction efficiency between the two species decreases significantly with increasing distance. This remains challenging for colloidal QDs, as they are generally covered with long alkyl chain ligands. For example, for efficient CT or Dexter exchange-type ET, which rely on donor-acceptor spatial wavefunction overlap, keeping the donor-acceptor distance below 1 nm is very important,[14-16] which is shorter than the commonly used ligands for QDs (such as oleic acid and oleylamine, around 2 nm in length).[17]

One common method for designing QD-molecular hybrids is directly mixing QDs with desired molecules that do not necessarily possess high binding affinity to the QD surfaces (**Scheme 1a**).[13, 18-20] This approach is limited for systems designed to interact via Förster resonance ET (FRET) but is not very efficient for CT or Dexter exchange-type ET due to inadequate minimum achievable donor-acceptor distance and low number of attached functional molecules. A more widespread way to design QD-molecular systems is by anchoring functional molecules with a binding group tailored to the QD surface (**Scheme 1b**), often via a ligand exchange procedure.[9-11, 16, 21-31] However, replacing too many native stabilizing ligands with small functional molecules can compromise QD's long-term colloidal stability. A better approach is developing ligands containing three key components: 1) strong binding to the QDs, 2) colloidal stability, and 3) the desired functionality (e.g., electron or hole scavenger molecule). This design, as shown in **Scheme 1c**, would result in a high loading of ligand tail groups for colloidal stability, a reduction of donor-acceptor distance, and high loading of the required functional group, allowing us to tune the number of acceptors a donor can interact with, thus increasing the probability of donor-acceptor interactions.

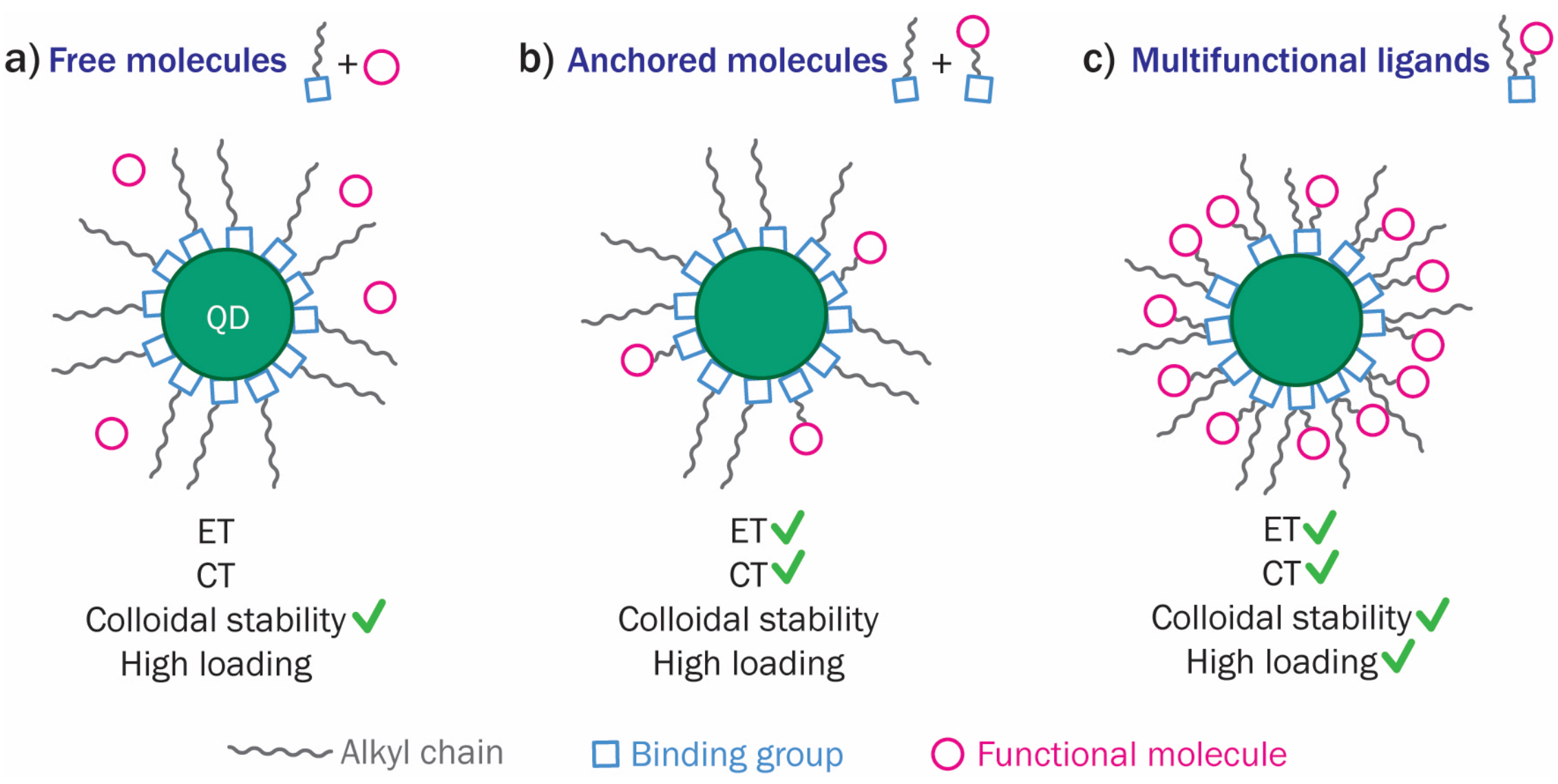

**Scheme 1. Designing QD-molecule hybrid.** A general schematic illustration of different approaches for designing a QD-molecule hybrid system for efficient ET or CT between them.

Perovskite QDs (pQDs) and nanocrystals have emerged as promising materials for optoelectronic devices,[32-34] and, more recently, also have been actively studied for photocatalycis.[5, 6, 28, 35, 36] Unlike conventional QDs, such as CdSe or InP, that usually require thick protective inorganic shell coatings with a large-band-gap material to passivate surface defects and an additional ligand shell to achieve colloidal stability, coating pQDs with ligands only is sufficient to passivate surface traps and achieve high quantum yields.[37] This defect tolerance of pQDs makes their interaction with molecular species significantly easier since the inorganic large-band-gap shell barrier is missing. This makes pQDs ideal candidates for designing efficient QD-molecular hybrids for photocatalysis,[6, 9] as for instance, has been demonstrated for various oxidation,[38] and bromination reactions.[39]

Some works rely on the dynamic interaction between molecules and pQD, with the molecules lacking affinity to pQD surfaces (see **Scheme 1a**).[18, 20, 40] Anchoring donor/acceptor molecules to pQDs should result in much more efficient transfer, however this route is relatively unexplored compared to conventional QDs. For example, several works use functional molecules with a special binding group with an affinity to pQD surfaces (see **Scheme 1b**).[11, 26, 27, 36, 41, 42] However, in the majority of cases dealing with CT, up to two orders of magnitude excess of molecular acceptors is needed to achieve high CT extraction efficiencies (from 53 up to 99%),[26, 28, 35, 43] which may undermine long-term pQD colloidal stability. In addition, the vast majority of authors use amine or carboxylic acid binding groups,[26, 27, 36, 42, 44-46] reminiscent of the chemistry of conventional QDs (e.g., anthracene-9-carboxylic acid, perylene diimide, ferrocenecarboxylic acid) .[21, 29, 38, 39, 47] These types of molecules have similar binding affinity to the pQDs as oleylamine or oleic acid, which is generally weak.[48] This means that such molecules can only be added in a post-synthesis step via ligand exchange, as they are readily lost during pQD purification, and often desorb over time. Additionally, these molecules provide no colloidal and chemical stability to pQDs, which limits their ability to achieve high loading on pQD surfaces for near-unity charge extraction, and thus, restricting their practical use.

In this work, we propose using multifunctional ligands for pQDs to make efficient pQD-molecule hybrids (see **Scheme 1c**). We engineer such ligands featuring a quaternary ammonium binding group with high affinity for the pQD surface, a long tail group for colloidal stability, and a functional group capable of electronically interacting with the pQDs. For the functional group, we use ferrocene, a well-known hole scavenger for both conventional QDs and perovskite NCs,[16, 31] and synthesize pQDs exclusively capped with these multifunctional ligands, resulting in stable pQDs. Using steady-state and time-resolved spectroscopy measurements, we find the hole transfer rate to be several picoseconds, with hole transfer efficiency up to 99.9%. We can tune the hole extraction efficiency over several orders of magnitude by capping the pQDs with different mixtures containing the multifunctional ferrocene ligands and didodecyldimethylammonium bromide (DDAB) ligands. Using density functional theory (DFT) calculations, we demonstrate that the ferrocene undergoes structure reorganization following CT, which affects its electronic orbitals and, ultimately, charge separation efficiency. Finally, to demonstrate the versatility of our approach, we synthesize pQDs with multifunctional ligands featuring nitrobenzene and anthracene functional groups, exhibiting efficient electron CT and ET, respectively. This facile approach of multifunctional ligand engineering for pQDs provides a blueprint for designing efficient QD-molecular hybrids for different optoelectronic applications, such as photocatalysis.

To functionalize $CsPbBr_3$ pQDs, we synthesized multifunctional ligands with a dimethylammonium bromide binding group, a long colloidal tail group, and an additional functional group *R* via the exothermic Menshutkin reaction (see **Materials and Methods** in **Supporting Information**).[49-51] Due to the availably of tertiary dimethylamines and alkyl halides, this reaction allows for the creation of a wide range of tailored ligands with varying colloidal tail group lengths and functional molecules. These multifunctional ligands are similar to DDAB ligands for pQDs but with one tail group replaced by a functional group *R* (**Figure 1a**, blue). The functional group can be selected to imprint a desired functionality onto the ligand. We choose the DDAB-like template for our ligands due to its strong binding affinity to the $CsPbBr_3$ surface and their ability to passivate surface defect states, leading to high PLQY in pQDs.[52] As the functional group, we choose ferrocene, a well-known hole scavenger for pQDs (**Figure 1**),[23, 24] resulting in (ferrocenylmethyl)dodecyldimethylammonium bromide (FeDDAB), with a yield of approximately 92% as determined by nuclear magnetic resonance (NMR) spectroscopy (see **Materials and Methods** in **Supporting Information**). Since the multifunctional ligand precursors do not interact with the pQDs, the synthesized ligands can be used directly without further purification.

**Figure 1. Synthesis of multifunctional ligands for pQDs. a)** Simplified schematic of the multifunctional ligands used in this work *(left)*, and a specific ligand with ferrocene as the functional group *R* and a dodecyl tail group for colloidal stability, resulting in the FeDDAB ligand *(right)*. **b)** Overview of the single-step Menshutkin reaction for synthesizing multifunctional ligands for pQDs.

For this study, we synthesized $CsPbBr_3$ pQDs with a diameter of roughly 6 nm using a slightly modified method described by Akkerman et al. (see **Materials and Methods** in **Supporting Information**).[53] As a reference, we prepared $CsPbBr_3$ pQDs capped exclusively with DDAB ligands. **Figure 2a** shows the absorbance, photoluminescence (PL, excited at 3.10 eV) spectra of DDAB-capped pQDs. The absorbance spectrum of these pQDs display several distinct features attributed to weakly confined excitons (labeled X1, X2).[54] As expected, DDAB-capped pQDs exhibit bright PL, with a PLQY of up to 80%. Next, we synthesized pQDs using only FeDDAB ligands. As shown in **Figure 2b**, the absorbance of the purified FeDDAB-capped pQDs is nearly identical to that of the DDAB-capped pQDs, indicating that even at 100% FeDDAB ligand concentration, and after purification, the pQDs remain monodisperse and colloidally stable (**Figure S1**).[53] In contrast to DDAB-capped pQDs, the FeDDAB-capped pQDs barely emit any light, with a PLQY of less than 0.1 %, close to the detection limit. We chose 3.10 eV (400 nm) as the excitation energy, which is near the absorbance minimum of FeDDAB (compare **Figure S2**), to predominantly excite the pQDs. The absorbance of FeDDAB-capped pQDs shows no detectable signal from FeDDAB ligands, which absorb light very weakly. The shape and energetic position of the spectra are independent of FeDDAB ligand concentration, pointing towards a weak coupling between pQD and ferrocene molecules (**Figure 2b, Figure S3** and **Figure S4**).[21, 22] Due to the multifunctional nature of FeDDAB ligands, the purified pQDs remain stable for at least 1 month, with no notable change in absorption (**Figure S5**). This further supports the strong binding of FeDDAB ligands to the $CsPbBr_3$ pQDs, and the lack of desorption of ferrocene from the pQDs. To confirm this, we performed NMR analysis on both the free ligands and FeDDAB-capped $CsPbBr_3$ pQDs (**Figure 2c, Figure S6,** and **Materials and Methods** in **Supporting Information**). As expected, FeDDAB–capped pQDs show significant broadening of the characteristic signal from the ferrocene group (between 4 and 5 ppm) as well as the methyl groups attached to the amine, indicating strong surface attachment of the FeDDAB ligands to pQDs.[55,56] Furthermore, the absence of sharp signals in the NMR spectrum indicative of free ligands confirms that a single purification step is sufficient to produce a clean pQD dispersion. Likewise, DFT calculation on the binding energy of the ligands,[11] confirm the similar binding energy values for the DDAB and FeDDAB ligands to the CsPbBr3 surfaces (40 and 42 kcal/mol respectively).

Due to the same binding groups of DDAB and FeDDAB ligands, we can precisely tune the ratio of FeDDAB to DDAB ligands on the pQDs. As depicted in **Figure 2d,** varying the percentage of FeDDAB in the total ligand amount during synthesis allows precise control over the pQD PLQY, spanning over several orders of magnitude (from 80% to <0.1%). The shape of the PL spectra of pQDs with different ligand ratios remains unchanged (**Figure S4**). The absorbance spectra of pQDs with the mixed ligand shells exhibit sharp excitonic resonances, similar to those of pQDs capped with 100% DDAB or 100% FeDDAB ligands (**Figure S3**).

To highlight the importance of designing multifunctional ligands tailored for the ionic surfaces of pQD, we performed a post-synthetic titration of ferrocene and FeDDAB on DDAB-capped QDs. As shown in **Figure 2e**, titrating ferrocene without the dimethylammonium group into a toluene solution of DDAB-capped pQDs does not affect the pQD PL intensity at low ferrocene concentrations, as ferrocene molecules do not bind to the ionic perovskite surface, and the layer of strongly bound DDAB ligands limits access to the pQD surface. Consequently, ferrocene molecules are too far away to engage in efficient excited-state interactions with pQDs, leaving their PL unaffected. This result is consistent with reports in literature.[35, 36] In contrast, titrating FeDDAB ligands to a toluene solution of DDAB-capped pQDs results in clear quenching of the pQD PL intensity, even at low FeDDAB molecule concentrations (**Figure 2e**). The direct synthesis and the post-synthetic exchange result in comparable PL quenching of the pQDs with similar FeDDAB and DDAB loading, indicating similar transfer efficiencies (see **Materials and Methods** in **Supporting Information** and **Figure S7**) and consequently, transfer rates. This demonstrates the importance of the donor-acceptor distance for efficient interaction, and emphasizes the need for rationally designed ligands tailored to the ionic surfaces of perovskites.

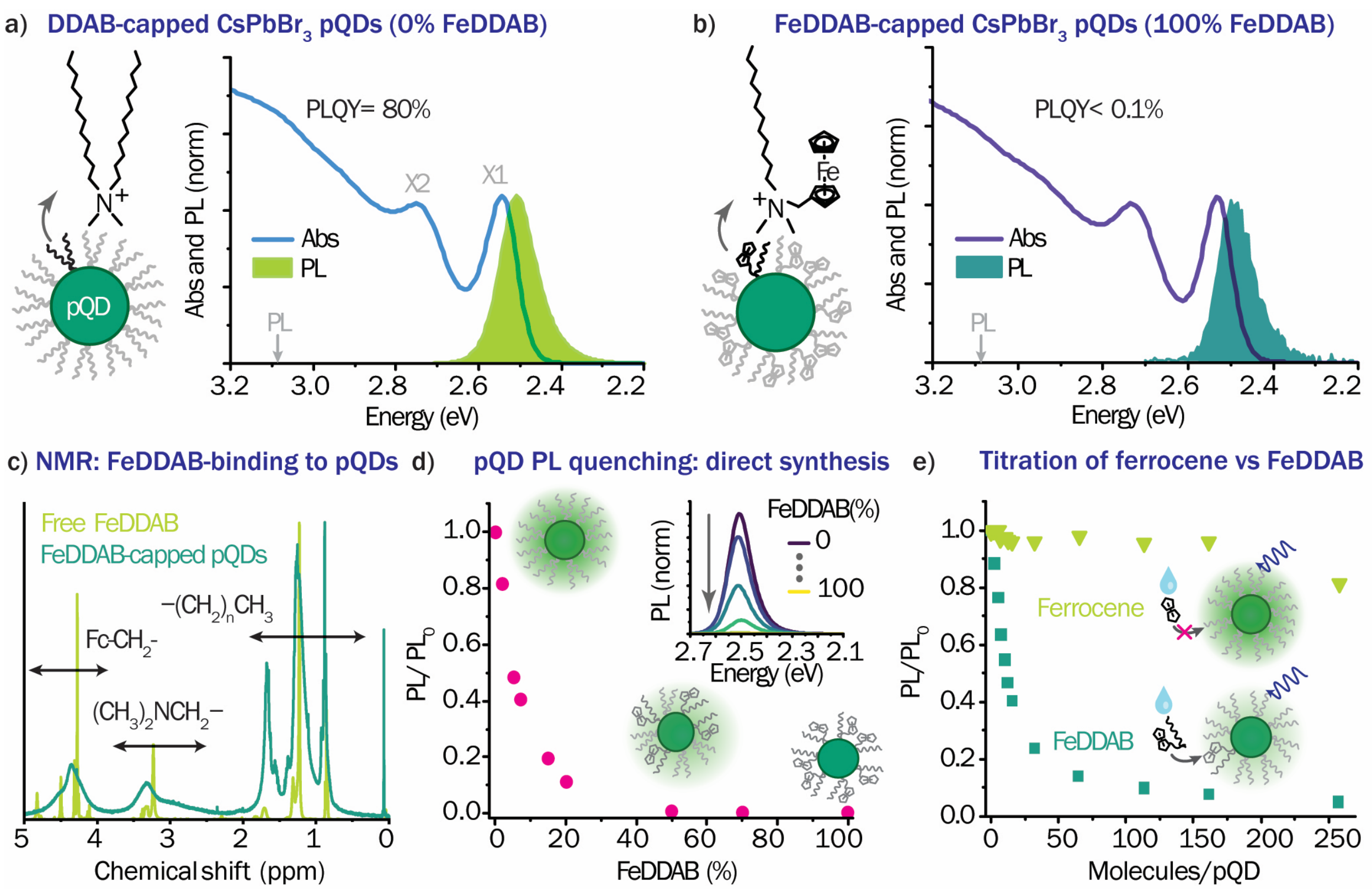

**Figure 2. Excited state interactions and binding between ferrocene functionalized ligands and**

**$CsPbBr_3$ pQDs.** Absorbance and PL spectra of **a)** DDAB- and **b)** FeDDAB-capped pQDs. **c)** NMR spectra of free FeDDAB ligands and FeDDAB-capped pQDs. In the latter, the FeDDAB signal shows significant broadening, indicative of strong binding to the pQD surface. **d)** Integrated PL intensity of pQDs synthesized with varying amounts of FeDDAB ligands, normalized to the integrated PL intensity of purely DDAB-capped pQDs. % refers to the percentage of FeDDAB ligands in the total ligand solution, which is a mixture of FeDDAB and DDAB ligands. Inset: PL spectra of the corresponding samples. **d)** Post-synthetic titration of ferrocene and FeDDAB molecules into a toluene solution containing DDAB-capped pQDs.

To investigate excited-state interactions between $CsPbBr_3$ pQDs and FeDDAB ligands, we performed time-resolved PL (TRPL) measurements using a streak camera. **Figures 3a** and **3b** show time-resolved 3D streak spectra of DDAB- and FeDDAB-capped pQDs, respectively, following selective excitation of pQDs at 3.10 eV. The TRPL profiles of both pQDs exhibit a single exponential profile. The DDAB-capped pQDs having a PL lifetime of $\kappa_{PL}^{-1} = 3.5$ ns (**Figure S8**). In contrast, the lifetime of FeDDAB-capped pQDs is shortened dramatically to 6.1 ps (**Figure 3c**), which is nearly three orders of magnitude faster compared to the PL lifetime of DDAB-capped pQDs. In addition, the maximum PL intensity is also reduced despite fixed excitation power, the same sample OD at the excitation and X1 energies, and a fixed experiment duration. Ferrocene has a type II energetic band alignment relative to the $CsPbBr_3$ pQDs, favoring photoexcited hole transfer from pQDs to ferrocene,[30, 36, 57] which might be the underlying reason behind pQD PL quenching.

To confirm that the rapid quenching of FeDDAB-capped pQD PL intensity is due to hole transfer from pQDs to ferrocene, we performed femtosecond transient absorption (TA) experiments (see **Materials and Methods** in **Supporting Information**). Samples were non-resonantly excited at 3.10 eV using ~100 fs pulses, and pump-induced changes in absorption ($\Delta A$) were monitored using a broad white light probe pulse. **Figure 3d** shows TA spectra of DDAB-capped pQDs. The negative feature at 2.55 eV can be assigned predominantly to the phase-space-filling of the energetically lowest exciton (X1 position).[54]

DDAB-capped pQDs exhibit a long-lived single-exponential $\Delta A$ signal at the X1 position, consistent with the TRPL lifetime (**Figure 3c**, light green dots; **Figure S9**). **Figure 3e** shows TA spectra of FeDDAB-capped pQDs. The shape of their TA spectra is comparable to the shape of the TA spectra of the DDAB-capped pQDs but have a significantly shorter lifetime. Notably, the $\Delta A$ signal at the X1 position of FeDDAB-capped pQDs recovers on slower timescales compared to their TRPL, and their maximum intensity is also reduced (compare **Figures 3c** and **3f**, dark green dots; **Figure S10a**).

The differing behavior of TRPL and $\Delta A$ signal of the energetically lowest exciton state in FeDDAB-capped pQDs provides clear evidence of hole transfer from pQDs to ferrocene molecules. In TRPL measurements, pQD PL is only detected if an exciton on a pQD recombines radiatively. If CT dissociates the exciton, radiative recombination of the dissociated electron-hole pair, emitting a photon of the same energy as the pristine exciton, is no longer possible. In TA experiments, however, the $\Delta A$ signal is proportional to the sum of the excited electron and hole population densities, where the presence of both charge carriers or either one of them on a pQD generates a signal, therefore, deviating from TRPL behavior during CT.[26, 27, 58-61] A positive $\Delta A$ signal around 2.00 eV, expected from charged ferrocene species,[35] is absent due to ferrocene's low absorbance (**Figure S11).** The nearly single-exponential and complete recovery of the $\Delta A$ signal at the energetic position of the X1 exciton on a timescale of 60 ps in FeDDAB-capped pQDs (**Figure S10b**) and the absence of low energetic CT exciton or exciplex emission,[62] suggest that hole transfer is soon followed by a subsequent electron transfer to the positively charge ferrocene.

To further support the claim that photoexcited hole transfer from a pQD to a ferrocene molecule is followed by electron transfer from the pQDs to the positively charged ferrocene, we model the $\Delta A$ signal of FeDDAB-capped pQDs, as discussed below. In a simplified view, the differential equation for the photoexcited hole population density $n_h$ in the measured pQD ensemble following their selective pulsed excitation can be approximated as $\dot{n_h}(t) \approx -\kappa_{ht} n_h(t)$ with the total hole transfer rate to the ferrocene molecule ensemble $\kappa_{ht} \gg \kappa_{PL}$. Here, we assume that ferrocene molecules remain in the ground state during the excitation of the pQDs. Given that the exciton Bohr diameter $2a_B$ is approximately 6 nm, comparable to the size of the pQDs, we assume that the exciton does not need to diffuse to the surface of the pQD to interact with a ferrocene molecule. Thus, $\kappa_{ht}$ describes the pure hole transfer rate. The differential equation for the electron population density $n_e$ in the measured pQD ensemble can be approximated as: $\dot{n}_e(t) \approx -(1 - e^{-\kappa_{ht}t})\kappa_{et}n_e(t)$ with the pure electron transfer rate $\kappa_{et} \gg \kappa_{PL}$. Electrons and holes in lead halide perovskites have comparable effective masses,[63] hence, the $\Delta A$ signal at the X1 position can be assumed to be proportional to the sum of electron and hole population densities in the measured pQD ensemble: $\Delta A \sim n_e + n_h$. Solving the differential equations, we find that $\Delta A = \alpha * [\exp(-\kappa_{ht}t) + exp\left(-\kappa_{et}t - \frac{\kappa_{et}}{\kappa_{ht}}(e^{\kappa_{ht}t} - 1)\right)]$ with a proportionality factor $\alpha$. Fitting this expression to the experimentally measured $\Delta A$ at the X1 position, assuming that $\kappa_{ht} = (6.1\ \text{ps})^{-1}$, as estimated from TRPL, and allowing $\alpha$ and $\kappa_{et}$ to vary, we obtain an electron transfer rate $\kappa_{et} = 0.067\ \text{ps}^{-1}$, equivalent to an electron transfer time of ~15 ps. This is about three times slower compared to the hole transfer time. Our results align with a previous report that positively charged ferrocene ions (ferrocenium) have a favorable band alignment with $CsPbBr_3$ pQD suitable for excited-state electron transfer from pQDs.[35] The hole transfer rate $\kappa_{ht}$ is a function of the number of ferrocene molecules attached to and interacting with a single pQD,[11, 16] whereas the electron can only transfer to the one photooxidized ferrocene attached to the same pQD.

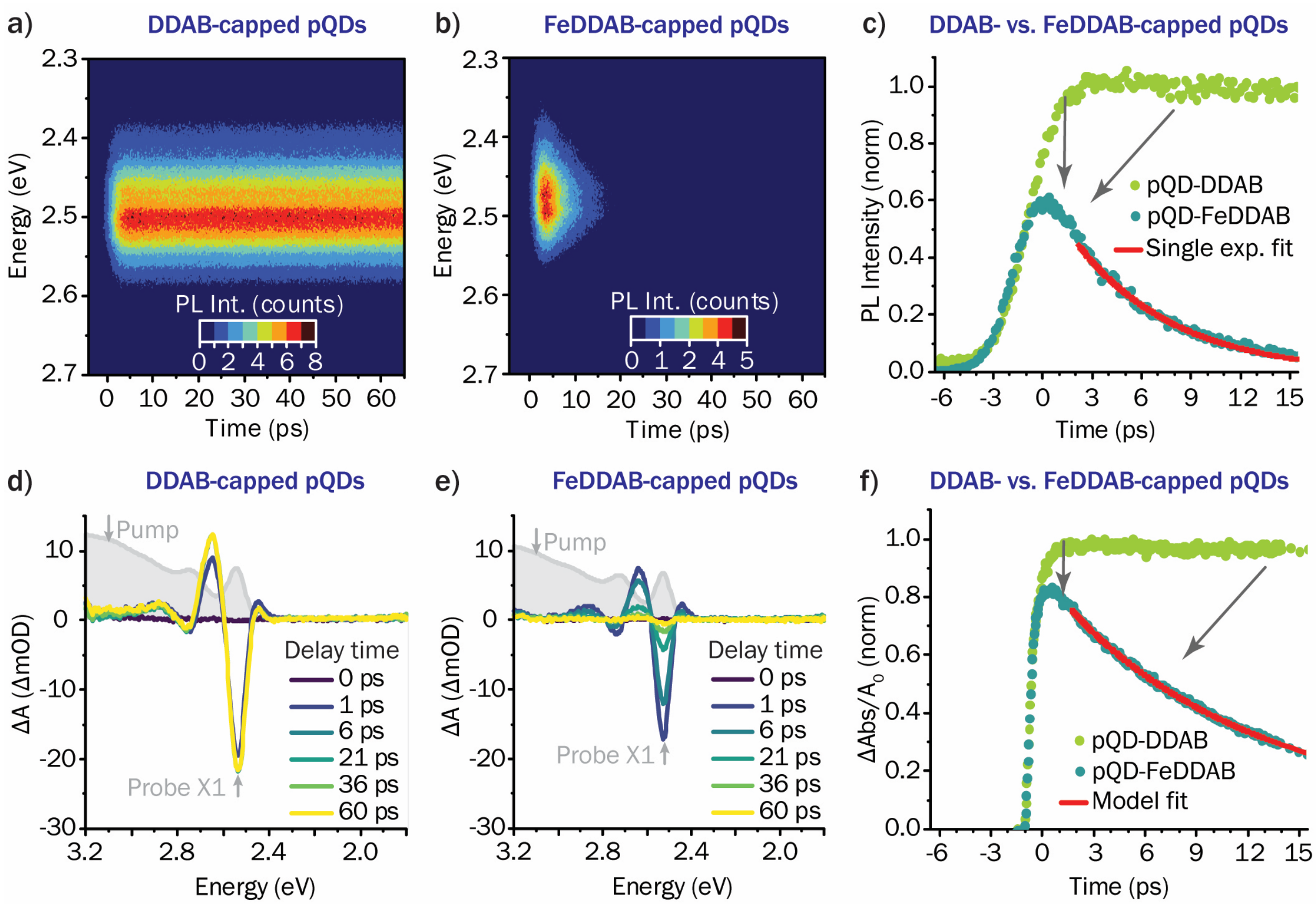

**Figure 3. Time-resolved PL and TA spectra of DDAB- and FeDDAB-capped $CsPbBr_3$ pQDs.** 3D streak spectra of **a)** DDAB- and **b)** FeDDAB-capped pQDs, excited at 3.10 eV. **c)** TRPL profiles of DDAB- (light green) and FeDDAB-capped (dark green) pQDs. Red line represents fit to the single exponential function ($R^2 = 0.991$). TA spectra of **d)** DDAB- and **e)** FeDDAB-capped pQDs, excited at 3.10 eV. Grey shaded areas represent steady-state absorbance spectra of the samples as guides for an eye. **f)** Comparison of the $\Delta A/A_0$ signal of X1 excitons in DDAB- (light green) and FeDDAB-capped (dark green) pQDs. $A_0$ is the steady-state OD of the sample at the X1 energy. Red line represents fit to the model function as discussed in the main text ($R^2 = 0.998$).

To confirm this hypothesis, we modeled a FeDDAB ligand attached to a 1.8-nm $CsPbBr_3$ pQD model (**Figure S12a**) at the DFT/Perdew-Burke-Ernzerhof (PBE) level of theory (see **Materials and**

**Methods** in **Supporting Information**). **Figure 4** illustrates the processes occurring following photoexcitation as computed with DFT by overlaying the electronic structure of the mixed pQD-ferrocene system. The calculations reveal an initial type-II energetic alignment with the FeDDAB highest occupied molecular orbital (HOMO) above the pQD valence band (VB). Upon photoexcitation, the pQD predominantly absorbs the incident light due its much larger absorption cross section, resulting in electron-hole pair generation, which is followed by the cooling of both charge carriers to the respective pQD band edges. The hole, however, is transferred to the ligand, as this process is energetically favored while the excited electron remains in the pQD conduction band (CB) edge. To predict the subsequent dynamics from this configuration, we calculated the lowest energy triplet state of the combined system. This state effectively mimics the lowest singlet excited state, apart from the spin-flip, since they share identical orbital occupation, and can be computed more straightforwardly using DFT. Immediately after charge transfer, the intermediate state, denoted as ($pQD^-$-$FeDDAB^+$), maintains the initial orbital occupation. However, spin symmetry breaking results in energy splitting between spin up and spin down orbitals, particularly evident in the frontier orbitals. Upon structural relaxation of this triplet state, $FeDDAB^+$ undergoes significant structural reorganization, with a notable elongation of the Fe-C bonds (**Figure S12b**). This elongation arises from the depopulation of the π-bonding Fe-C orbitals upon oxidation via hole transfer to the cationic state. This structural relaxation lowers the lowest unoccupied molecular orbital (LUMO) of the ferrocenium below the pQD CB edge, establishing a reverse type-I alignment that promotes ferrocenium reduction via electron transfer from the pQD. The newly formed (pQD-FeDDAB*) likely undergoes non-radiative electron-hole recombination within the FeDDAB ligand, finally restoring the (pQD-FeDDAB) ground state configuration, consistent with our time-resolved experiments and previous reports in literature.[35]

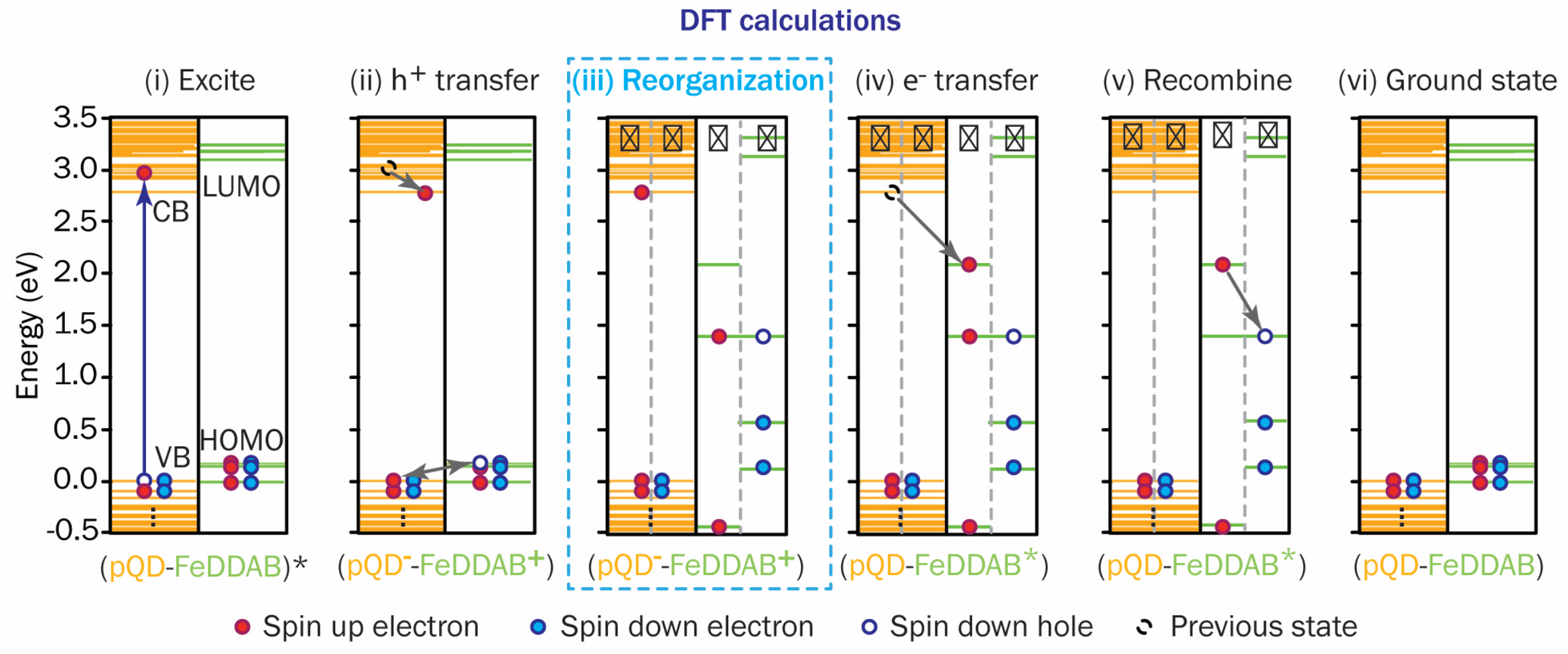

**Figure 4. DFT calculations of excited state interaction between pQD and ferrocene.** Energy band alignment computed at the DFT/PBE level of theory for a $CsPbBr_3$ pQD (orange) capped with a FeDDAB ligand molecule (green). The pQD VB is set as the energy origin. To better describe the carriers within the excited systems, the spin configurations - spin-up and spin-down, represented by ↑ and ↓, respectively - are explicitly shown. If not specified, spin-up and spin-down are considered energetically degenerate.

Based on these results, we can summarize the excited-state interaction between closely bound ferrocene molecules and pQDs as follows. Due to pQDs and ferrocene molecules being separated by a single carbon bond, the distance between them is very small, facilitating their spatial wavefunction overlap. As a result, upon photoexcitation of pQDs, the pQD-FeDDAB system can be understood as the entire complex being in an excited state: (pQD*-FeDDAB)→ (pQD-FeDDAB)*, as illustrated in **Scheme 2**. Due to the large energetic driving force for a hole to localize on a ferrocene molecule, charge separation happens on an average time scale of 6 ps, leading to the formation of a negatively charged pQDs and a positively charged ferrocene complex ($pQD^-$-$FeDDAB^+$). This corresponds to a hole transfer efficiency of $\frac{\kappa_{ht}}{\kappa_{PL}+\kappa_{ht}} \cdot 100\% = 99.9\%$,[16] which matches the PL quenching efficiency defined as $1 - \frac{PLQY_{pQD-FeDDAB}}{PLQY_{pQD-DDAB}} \cdot 100\% =$ 99.8%. Structural reorganization of ferrocene as a result of positive charge acquisition to minimize its energy alters its molecular orbital structure, creating a large driving force for electrons to localize on the ferrocenium molecule (pQD-FeDDAB*). Electron transfer happens on an average timescale of 15 ps and is ultimately followed by non-radiative recombination of excited electron-hole pairs on the ferrocene molecule. All in all, this results in a sequential electron and hole transfer to ferrocene. The importance of the donor-acceptor distance is emphasized in steady-state PL measurements shown in **Figure 2e**, where ferrocene molecules engage in efficient excited-state interactions with pQD only when part of the multifunctional ligands. Our work also highlights the importance of molecular structure reorganization following charge transfer, that affects the energetic position of molecular orbitals of the molecule and influences charge separation efficiency.

**Scheme 2. State picture illustrating charge transfer.** Schematic illustration of the excited state interaction between FeDDAB-capped pQDs and functionalized ferrocene molecules.

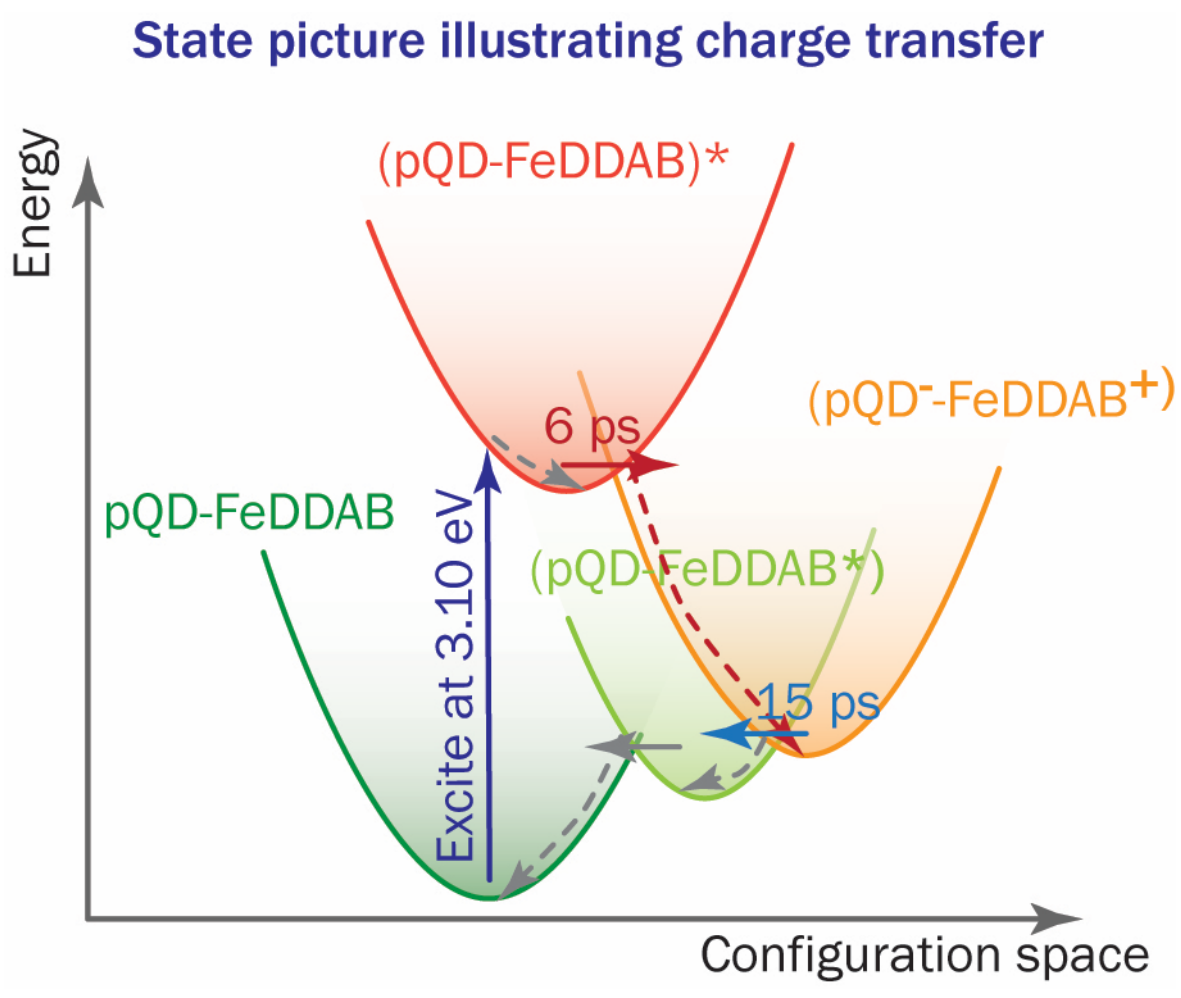

To extend our approach to also include electron CT, we synthesized a multifunctional ligand with a known electron-withdrawing group, nitrobenzyl, resulting in the (nitrobenzene) octadecyldimethylammonium bromide (NBODAB) (**Figure S13**). Similar to FeDDAB, the single-step Menshutkin reaction results in a high purity NBODAB, with a reaction yield of 99%, and no further purification is required. The NMR spectra of the NBODAB-capped pQDs show the characteristic signal of nitrobenzene protons around 8-9 ppm, which appear significantly broadened, confirm strong binding of the NBODAB ligands to the pQD surfaces **(Figure S14)**, as also supported with a similar DFT binding energy value of 44 kcal/mol compared to the 40 kcal/mol for DDAB.[11] As shown in **Figure 5a**, the absorbance spectrum of NBODAB-capped pQDs closely resembles the absorbance of DDAB-capped pQD. Similar to FeDDAB-capped pQDs, NBODAB-capped pQDs exhibit a significant PL quenching in comparison to DDAB-capped QDs (**Figure 5b**). These NBODAB-capped pQDs exhibit a PLQY of 0.8 %, suggesting an electron transfer efficiency of 99 % (relative to the DDAB caped QDs). To confirm the electron CT from the NBODAB ligands, we performed DFT calculations on NBODAB-capped pQDs (**Figure S12a**). These calculations reveal a type-II energetic alignment, where the LUMO of the NBODAB molecule is below the pQD CB, while its HOMO remains well within the pQD VB (**Figure 5c**). With this ground state electronic structure of the combined system, we expect that upon photoexcitation of the pQD, the excited electron initially cools to the CB edge and then is transferred to the ligand. Therefore, the combination of such fast and efficient hole and electron transfer from pQD to the tailored FeDDAB and NBODAB ligands highlight the potential of using pQDs in photocatalysis, both for oxidative as well as reductive reactions.

The use of multifunctional ligands is not limited to CT but can also enable ET. By selecting an appropriate functional group, such as anthracene, we synthesized anthracene dodecyl dimethyl ammonium bromide (ADDAB) obtained with 99% purity from the single-step Menshutkin reaction (**Figure S15a**). In contrast to the pQD PL quenching observed in FeDDAB- and NBODAB-capped pQDs, ADDAB-capped pQDs exhibit PL enhancement compared to DDAB-capped pQDs, suggesting ET from the anthracene functionalized ligands to the pQDs (**Figure S12b**).[45] This is further supported by PL excitation (PLE) measurements (**Figure S12c-d**). As with the other ligands discussed in this work, the NMR spectra of ADDAB-capped pQDs show significant broadening of the characteristic signal of anthracene protons around 7-9 ppm, confirming the strong binding of the ligands to the pQD surface **(Figure S16)**, as also supported with a similar DFT binding energy values of 50 kcal/mol compared to the 40 kcal/mol for DDAB.[11] These findings demonstrate the wide applicability of the multifunctional ligands, ranging from photocatalysis (via CT) to optoelectronics (via ET from multifunctional ligands to pQDs) and photovoltaics and sensing (ET from pQDs to ligand shell).

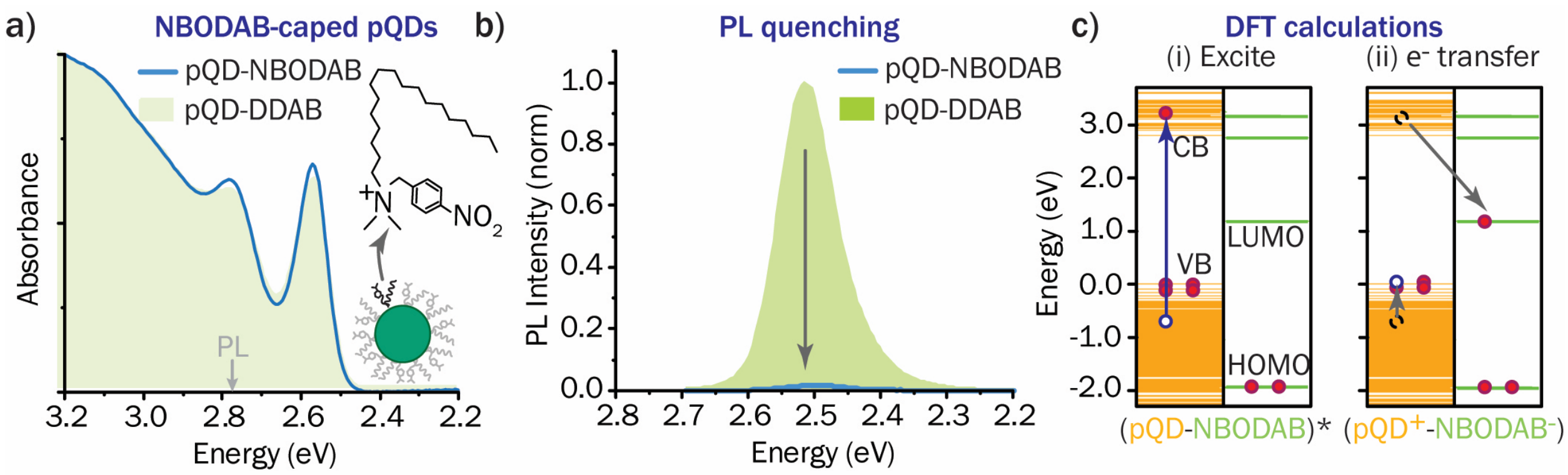

**Figure 5. Excited state interactions between multifunctional ligand functional groups and $CsPbBr_3$ pQDs. a)** Absorbance and **b)** PL spectra of NBODAB-capped pQDs (solid blue lines) and DDAB-capped pQDs (green shaded areas). **c)** Energy band alignment for the pQD-NBODAB system computed at the DFT/PBE level of theory $CsPbBr_3$ pQD model (orange) and a NBODAB ligand molecule (green) to illustrate carrier dynamics in the presence of the NBODAB ligand.

In summary, we showed how to design efficient pQD-molecular hybrids by engineering multifunctional ligands via the Menshutkin reaction. These ligands feature a quaternary ammonium binding group with high affinity to the pQD surface, a long tail group for colloidal stability, and a functional group capable of interacting with the pQDs. To demonstrate that our approach works, we synthesized pQDs exclusively capped with multifunctional FeDDAB ligands featuring ferrocene as the functional group. Using steady-state spectroscopy experiments, we showed that ferrocene molecules engage in efficient excited-state interactions with pQD only when part of the multifunctional ligands, emphasizing the importance of donor-acceptor distance for efficient interaction. We also demonstrated that the interaction efficiency between pQDs and ferrocene molecules can be tuned across several orders of magnitude by adjusting the ratio of DDAB and FeDDAB capping ligands. Using time-resolved spectroscopy measurements and DFT calculations, we provided evidence for photoexcited hole transfer from pQDs to ferrocene molecules, achieving a hole extraction rate as fast as 0.16 $ps^{-1}$, corresponding to a near-unity hole transfer efficiencyStructural reorganization of ferrocene as a result of positive charge acquisition alters its electronic band structure, creating a large driving force for electrons to localize on the ferrocenium molecule. Electron transfer happens on an average timescale of 15 ps and is ultimately followed by non-radiative recombination of excited electron-hole pairs on the ferrocene molecule. DFT calculations confirms this picture by demonstrating a significant structural reorganization of the FeDDAB when it receives the hole from the pQD. To show the versatility of our approach, we synthesized multifunctional ligands with nitrobenzene and anthracene functional groups, capable of engaging in photoexcited electron and energy transfer processes with pQDs, respectively. We also demonstrated that pQDs capped with multifunctional ligands remain colloidally stable for months, with the interaction dynamics between pQDs and functional molecules remaining unaltered. This facile approach of multifunctional ligand engineering for pQDs provides a blueprint for designing efficient QD-molecular hybrids for different photocatalytic and optoelectronic applications.

**Supporting Information**

The Supporting Information is available free of charge.

Materials and methods, NMR characterization of ligands and pQDs, additional data regarding binding of the ligands, TEM images, additional steady-state absorption and PL data, additional TRPL data, a comparison between the direct ligand attachment and the post-synthetic ligand exchange PL quenching, additional TAS measurements $CsPbBr_3$ pQD models used for DFT and pQDs capped with anthracene ligands.

**Notes**

The authors declare no competing financial interest.

**Acknowledgements**

We acknowledge financial support by the Bavarian State Ministry of Science, Research, and Arts and the LMU Munich through the grant "Solar Technologies go Hybrid (SolTech)". Q.A.A. acknowledges the LMUexcellent, funded by the Federal Ministry of Education and Research (BMBF) and the Free State of Bavaria under the Excellence Strategy of the Federal Government and the Länder. M.K and J.F. acknowledge support by a Grant from the GIF, the German-Israeli Foundation for Scientific Research and Development No. I-1512-401.10. We thank the Rädler Chair for providing access to the electron microscope. We also thank local research centers such as the Center of Nanoscience (CeNS) for providing communicative networking structures. II and JL acknowledge Horizon Europe EIC Pathfinder program through project 101098649─UNICORN. DFT calculations were carried at the Donostia International Physics (DIPC) Supercomputing Center, for which the authors acknowledge for the technical and human support. We also thank PRACE for awarding us access to Leonardo at CINECA, Italy.

**References**

1. Bae, W. K.; Brovelli, S.; Klimov, V. I., Spectroscopic insights into the performance of quantum dot light-emitting diodes. *Mrs Bulletin* **2013,** *38* (9), 721-730.
2. Lim, J.; Park, Y. S.; Wu, K.; Yun, H. J.; Klimov, V. I., Droop-Free Colloidal Quantum Dot Light-Emitting Diodes. *Nano Lett* **2018,** *18* (10), 6645-6653.
3. Gur, I.; Fromer, N. A.; Geier, M. L.; Alivisatos, A. P., Air-stable all-inorganic nanocrystal solar cells processed from solution. *Science* **2005,** *310* (5747), 462-5.
4. Kagan, C. R.; Bassett, L. C.; Murray, C. B.; Thompson, S. M., Colloidal Quantum Dots as Platforms for Quantum Information Science. *Chem. Rev.* **2021,** *121* (5), 3186-3233.
5. Wu, H. L.; Li, X. B.; Tung, C. H.; Wu, L. Z., Semiconductor Quantum Dots: An Emerging Candidate for CO(2) Photoreduction. *Adv Mater* **2019,** *31* (36), e1900709.
6. DuBose, J. T.; Kamat, P. V., Efficacy of Perovskite Photocatalysis: Challenges to Overcome. *ACS Energy Letters* **2022,** *7* (6), 1994-2011.
7. Choi, H.; Nicolaescu, R.; Paek, S.; Ko, J.; Kamat, P. V., Supersensitization of CdS quantum dots with a near-infrared organic dye: toward the design of panchromatic hybrid-sensitized solar cells. *ACS Nano* **2011,** *5* (11), 9238-45.
8. Wolff, C. M.; Frischmann, P. D.; Schulze, M.; Bohn, B. J.; Wein, R.; Livadas, P.; Carlson, M. T.; Jäckel, F.; Feldmann, J.; Würthner, F.; Stolarczyk, J. K., All-in-one visible-light-driven water splitting by combining nanoparticulate and molecular co-catalysts on CdS nanorods. *Nature Energy* **2018,** *3* (10), 862-869.

9. Gupta, S.; Singh, S.; De, S.; Gautam, N.; Patel, H.; Govind Rao, V., Perovskite–Molecular Photocatalyst Synergy and Surface Engineering for Superior Photocatalytic Performance. *ACS Applied Materials & Interfaces* **2025,** *17* (8), 12054-12063.
10. Singh, S.; Ganguly, D.; Gupta, S.; Govind Rao, V., Enhancing the Photocatalytic Performance of CsPbBr3 Nanocrystals through Ferrocene-Assisted Exciton Dissociation and Halide Vacancies. *ACS Applied Materials & Interfaces* **2024,** *16* (49), 67854-67861.
11. Kurashvili, M.; Llusar, J.; Stickel, L. S.; Wurthner, T.; Ederle, D.; Infante, I.; Feldmann, J.; Akkerman, Q. A., Efficient Energy Transfer from Quantum Dots to Closely-Bound Dye Molecules without Spectral Overlap. *Angew Chem Int Ed Engl* **2025,** *64* (9), e202420658.
12. Clapp, A. R.; Medintz, I. L.; Mauro, J. M.; Fisher, B. R.; Bawendi, M. G.; Mattoussi, H., Fluorescence resonance energy transfer between quantum dot donors and dye-labeled protein acceptors. *J Am Chem Soc* **2004,** *126* (1), 301-10.
13. Soujon, D.; Becker, K.; Rogach, A. L.; Feldmann, J.; Weller, H.; Talapin, D. V.; Lupton, J. M., Time-Resolved Förster Energy Transfer from Individual Semiconductor Nanoantennae to Single Dye Molecules. *The Journal of Physical Chemistry C* **2007,** *111* (31), 11511-11515.
14. Dexter, D. L., A Theory of Sensitized Luminescence in Solids. *The Journal of Chemical Physics* **1953,** *21* (5), 836-850.
15. Govorov, A.; Hernández Martínez, P. L.; V., D. H., Understanding and Modeling Förster-type Resonance Energy Transfer (FRET). *SpringerBriefs in Applied Sciences and Technology* **2016**.
16. Ding, T. X.; Olshansky, J. H.; Leone, S. R.; Alivisatos, A. P., Efficiency of hole transfer from photoexcited quantum dots to covalently linked molecular species. *J Am Chem Soc* **2015,** *137* (5), 2021-9.
17. Mourdikoudis, S.; Menelaou, M.; Fiuza-Maneiro, N.; Zheng, G.; Wei, S.; Perez-Juste, J.; Polavarapu, L.; Sofer, Z., Oleic acid/oleylamine ligand pair: a versatile combination in the synthesis of colloidal nanoparticles. *Nanoscale Horiz* **2022,** *7* (9), 941-1015.
18. Feld, L. G.; Boehme, S. C.; Morad, V.; Sahin, Y.; Kaul, C. J.; Dirin, D. N.; Raino, G.; Kovalenko, M. V., Quantifying Forster Resonance Energy Transfer from Single Perovskite Quantum Dots to Organic Dyes. *ACS Nano* **2024,** *18* (14), 9997-10007.
19. Becker, K.; Lupton, J. M.; Muller, J.; Rogach, A. L.; Talapin, D. V.; Weller, H.; Feldmann, J., Electrical control of Forster energy transfer. *Nat Mater* **2006,** *5* (10), 777-81.
20. Wu, K.; Liang, G.; Shang, Q.; Ren, Y.; Kong, D.; Lian, T., Ultrafast Interfacial Electron and Hole Transfer from CsPbBr3 Perovskite Quantum Dots. *J Am Chem Soc* **2015,** *137* (40), 12792-5.
21. Lian, S.; Weinberg, D. J.; Harris, R. D.; Kodaimati, M. S.; Weiss, E. A., Subpicosecond Photoinduced Hole Transfer from a CdS Quantum Dot to a Molecular Acceptor Bound Through an Exciton-Delocalizing Ligand. *ACS Nano* **2016,** *10* (6), 6372-82.
22. Wang, K.; Cline, R. P.; Schwan, J.; Strain, J. M.; Roberts, S. T.; Mangolini, L.; Eaves, J. D.; Tang, M. L., Efficient photon upconversion enabled by strong coupling between silicon quantum dots and anthracene. *Nat Chem* **2023,** *15* (8), 1172-1178.
23. Dorokhin, D.; Tomczak, N.; Velders, A. H.; Reinhoudt, D. N.; Vancso, G. J., Photoluminescence Quenching of CdSe/ZnS Quantum Dots by Molecular Ferrocene and Ferrocenyl Thiol Ligands. *The Journal of Physical Chemistry C* **2009,** *113* (43), 18676-18680.
24. Tarafder, K.; Surendranath, Y.; Olshansky, J. H.; Alivisatos, A. P.; Wang, L. W., Hole transfer dynamics from a CdSe/CdS quantum rod to a tethered ferrocene derivative. *J Am Chem Soc* **2014,** *136* (13), 5121-31.
25. Olshansky, J. H.; Ding, T. X.; Lee, Y. V.; Leone, S. R.; Alivisatos, A. P., Hole Transfer from Photoexcited Quantum Dots: The Relationship between Driving Force and Rate. *J Am Chem Soc* **2015,** *137* (49), 15567-75.

26. Gelvez-Rueda, M. C.; Fridriksson, M. B.; Dubey, R. K.; Jager, W. F.; van der Stam, W.; Grozema, F. C., Overcoming the exciton binding energy in two-dimensional perovskite nanoplatelets by attachment of conjugated organic chromophores. *Nat Commun* **2020,** *11* (1), 1901.
27. Wei, Z.; Mulder, J. T.; Dubey, R. K.; Evers, W. H.; Jager, W. F.; Houtepen, A. J.; Grozema, F. C., Tuning the Driving Force for Charge Transfer in Perovskite-Chromophore Systems. *J Phys Chem C Nanomater Interfaces* **2023,** *127* (31), 15406-15415.
28. Kobosko, S. M.; DuBose, J. T.; Kamat, P. V., Perovskite Photocatalysis. Methyl Viologen Induces Unusually Long-Lived Charge Carrier Separation in CsPbBr3 Nanocrystals. *ACS Energy Letters* **2019,** *5* (1), 221-223.
29. Morris-Cohen, A. J.; Peterson, M. D.; Frederick, M. T.; Kamm, J. M.; Weiss, E. A., Evidence for a Through-Space Pathway for Electron Transfer from Quantum Dots to Carboxylate-Functionalized Viologens. *The Journal of Physical Chemistry Letters* **2012,** *3* (19), 2840-2844.
30. Singh, S.; Mittal, D.; Gurunarayanan, V.; Sahu, A.; Ramapanicker, R.; Govind Rao, V., Binding Strength-Guided Shuttling of Charge Carriers from Perovskite Nanocrystals to Molecular Acceptors. *ACS Applied Energy Materials* **2023,** *6* (15), 8091-8101.
31. De, S.; Singh, S.; Aggarwal, P.; Govind Rao, V., Mapping Binding Sites for Efficient Hole Extraction in Lead Halide Perovskites through Sulfur-Based Ligand Engineering. *Advanced Optical Materials* **2025,** *13* (6), 2402562.
32. Hu, L.; Zhao, Q.; Huang, S.; Zheng, J.; Guan, X.; Patterson, R.; Kim, J.; Shi, L.; Lin, C. H.; Lei, Q.; Chu, D.; Tao, W.; Cheong, S.; Tilley, R. D.; Ho-Baillie, A. W. Y.; Luther, J. M.; Yuan, J.; Wu, T., Flexible and efficient perovskite quantum dot solar cells via hybrid interfacial architecture. *Nat Commun* **2021,** *12* (1), 466.
33. Zhao, Q.; Hazarika, A.; Chen, X.; Harvey, S. P.; Larson, B. W.; Teeter, G. R.; Liu, J.; Song, T.; Xiao, C.; Shaw, L.; Zhang, M.; Li, G.; Beard, M. C.; Luther, J. M., High efficiency perovskite quantum dot solar cells with charge separating heterostructure. *Nat Commun* **2019,** *10* (1), 2842.
34. Wang, Y.; Yuan, J.; Zhang, X.; Ling, X.; Larson, B. W.; Zhao, Q.; Yang, Y.; Shi, Y.; Luther, J. M.; Ma, W., Surface Ligand Management Aided by a Secondary Amine Enables Increased Synthesis Yield of CsPbI3 Perovskite Quantum Dots and High Photovoltaic Performance. *Advanced Materials* **2020,** *32* (32), 2000449.
35. DuBose, J. T.; Kamat, P. V., Probing Perovskite Photocatalysis. Interfacial Electron Transfer between CsPbBr(3) and Ferrocene Redox Couple. *J Phys Chem Lett* **2019,** *10* (20), 6074-6080.
36. Singh, S.; Mittal, D.; Gurunarayanan, V.; Sahu, A.; Ramapanicker, R.; Govind Rao, V., Perovskite photocatalysis: realizing long-lived charge-separated states at the interface of CsPbBr3nanocrystals and functionalized ferrocene molecules. *Journal of Materials Chemistry A* **2022,** *10* (39), 21112-21123.
37. Nedelcu, G.; Protesescu, L.; Yakunin, S.; Bodnarchuk, M. I.; Grotevent, M. J.; Kovalenko, M. V., Fast Anion-Exchange in Highly Luminescent Nanocrystals of Cesium Lead Halide Perovskites (CsPbX3, X = Cl, Br, I). *Nano Lett* **2015,** *15* (8), 5635-40.
38. Pradhan, S.; Bhujel, D.; Gurung, B.; Sharma, D.; Basel, S.; Rasaily, S.; Thapa, S.; Borthakur, S.; Ling, W. L.; Saikia, L.; Reiss, P.; Pariyar, A.; Tamang, S., Stable lead-halide perovskite quantum dots as efficient visible light photocatalysts for organic transformations. *Nanoscale Advances* **2021,** *3* (5), 1464-1472.
39. Amberg, W. M.; Lindner, H.; Sahin, Y.; Staudinger, E.; Morad, V.; Sabisch, S.; Feld, L. G.; Li, Y.; Dirin, D. N.; Kovalenko, M. V.; Carreira, E. M., Ligand Influence on the Performance of Cesium Lead Bromide Perovskite Quantum Dots in Photocatalytic C(sp3)–H Bromination Reactions. *Journal of the American Chemical Society* **2025,** *147* (10), 8548-8558.
40. De, A.; Das, S.; Samanta, A., Hot Hole Transfer Dynamics from CsPbBr3 Perovskite Nanocrystals. *ACS Energy Letters* **2020,** *5* (7), 2246-2252.

41. Chakkamalayath, J.; Martin, L. E.; Kamat, P. V., Extending Infrared Emission via Energy Transfer in a CsPbI(3)-Cyanine Dye Hybrid. *J Phys Chem Lett* **2024,** *15* (2), 401-407.
42. Luo, X.; Han, Y.; Chen, Z.; Li, Y.; Liang, G.; Liu, X.; Ding, T.; Nie, C.; Wang, M.; Castellano, F. N.; Wu, K., Mechanisms of triplet energy transfer across the inorganic nanocrystal/organic molecule interface. *Nat Commun* **2020,** *11* (1), 28.
43. DuBose, J. T.; Kamat, P. V., Surface Chemistry Matters. How Ligands Influence Excited State Interactions between CsPbBr3 and Methyl Viologen. *The Journal of Physical Chemistry C* **2020,** *124* (24), 12990-12998.
44. Rossi, A.; Price, M. B.; Hardy, J.; Gorman, J.; Schmidt, T. W.; Davis, N. J. L. K., Energy Transfer between Perylene Diimide Based Ligands and Cesium Lead Bromide Perovskite Nanocrystals. *The Journal of Physical Chemistry C* **2020,** *124* (5), 3306-3313.
45. Hardy, J.; Brett, M. W.; Rossi, A.; Wagner, I.; Chen, K.; Timmer, M. S. M.; Stocker, B. L.; Price, M. B.; Davis, N. J. L. K., Energy Transfer between Anthracene-9-carboxylic Acid Ligands and CsPbBr3 and CsPbI3 Nanocrystals. *The Journal of Physical Chemistry C* **2021,** *125* (2), 1447-1453.
46. He, S.; Lai, R.; Jiang, Q.; Han, Y.; Luo, X.; Tian, Y.; Liu, X.; Wu, K., Engineering Sensitized Photon Upconversion Efficiency via Nanocrystal Wavefunction and Molecular Geometry. *Angew Chem Int Ed Engl* **2020,** *59* (40), 17726-17731.
47. Lai, R.; Liu, Y.; Luo, X.; Chen, L.; Han, Y.; Lv, M.; Liang, G.; Chen, J.; Zhang, C.; Di, D.; Scholes, G. D.; Castellano, F. N.; Wu, K., Shallow distance-dependent triplet energy migration mediated by endothermic charge-transfer. *Nat Commun* **2021,** *12* (1), 1532.
48. De Roo, J.; Ibanez, M.; Geiregat, P.; Nedelcu, G.; Walravens, W.; Maes, J.; Martins, J. C.; Van Driessche, I.; Kovalenko, M. V.; Hens, Z., Highly Dynamic Ligand Binding and Light Absorption Coefficient of Cesium Lead Bromide Perovskite Nanocrystals. *ACS Nano* **2016,** *10* (2), 2071-81.
49. Menschutkin, N., Über die Affinitätskoeffizienten der Alkylhaloide und der Amine. *Zeitschrift für Physikalische Chemie* **1890,** *6U* (1), 41-57.
50. Menschutkin, N., Beiträge zur Kenntnis der Affinitätskoeffizienten der Alkylhaloide und der organischen Amine. *Zeitschrift für Physikalische Chemie* **1890,** *5U* (1), 589-600.
51. Abboud, J. M.; Notario, R.; Bertrán, J.; Solà, M., One Century of Physical Organic Chemistry: The Menshutkin Reaction. In *Progress in Physical Organic Chemistry*, 1993; pp 1-182.
52. Bodnarchuk, M. I.; Boehme, S. C.; Ten Brinck, S.; Bernasconi, C.; Shynkarenko, Y.; Krieg, F.; Widmer, R.; Aeschlimann, B.; Gunther, D.; Kovalenko, M. V.; Infante, I., Rationalizing and Controlling the Surface Structure and Electronic Passivation of Cesium Lead Halide Nanocrystals. *ACS Energy Lett* **2019,** *4* (1), 63-74.
53. Akkerman, Q. A., Spheroidal Cesium Lead Chloride-Bromide Quantum Dots and a Fast Determination of Their Size and Halide Content. *Nano Lett* **2022,** *22* (20), 8168-8173.
54. Barfusser, A.; Rieger, S.; Dey, A.; Tosun, A.; Akkerman, Q. A.; Debnath, T.; Feldmann, J., Confined Excitons in Spherical-Like Halide Perovskite Quantum Dots. *Nano Lett* **2022,** *22* (22), 8810-8817.
55. Hens, Z.; Martins, J. C., A Solution NMR Toolbox for Characterizing the Surface Chemistry of Colloidal Nanocrystals. *Chemistry of Materials* **2013,** *25* (8), 1211-1221.
56. Hens, Z., Ligands on Nanocrystal Surfaces, the 1H Nuclear Magnetic Resonance Fingerprint. *Accounts of Chemical Research* **2023,** *56* (12), 1623-1633.
57. Ahlawat, M.; Kumari, S.; Govind Rao, V., Synergistic binding between an engineered interface and functionalized ferrocene offers remarkable charge extraction efficiency in lead halide perovskites. *Journal of Materials Chemistry A* **2023,** *11* (25), 13289-13299.
58. Richter, A. F.; Binder, M.; Bohn, B. J.; Grumbach, N.; Neyshtadt, S.; Urban, A. S.; Feldmann, J., Fast Electron and Slow Hole Relaxation in InP-Based Colloidal Quantum Dots. *ACS Nano* **2019,** *13* (12), 14408-14415.

59. Luo, X.; Liang, G.; Han, Y.; Li, Y.; Ding, T.; He, S.; Liu, X.; Wu, K., Triplet Energy Transfer from Perovskite Nanocrystals Mediated by Electron Transfer. *J Am Chem Soc* **2020,** *142* (25), 11270-11278.
60. Yao, E. P.; Bohn, B. J.; Tong, Y.; Huang, H.; Polavarapu, L.; Feldmann, J., Exciton Diffusion Lengths and Dissociation Rates in CsPbBr3 Nanocrystal–Fullerene Composites: Layer-by-Layer versus Blend Structures. *Advanced Optical Materials* **2019,** *7* (8), 1801776.
61. Dey, A.; Strohmair, S.; He, F.; Akkerman, Q. A.; Feldmann, J., Fast electron and slow hole spin relaxation in CsPbI3 nanocrystals. *Applied Physics Letters* **2022,** *121* (20), 201106.
62. Deschler, F.; Da Como, E.; Limmer, T.; Tautz, R.; Godde, T.; Bayer, M.; von Hauff, E.; Yilmaz, S.; Allard, S.; Scherf, U.; Feldmann, J., Reduced charge transfer exciton recombination in organic semiconductor heterojunctions by molecular doping. *Phys Rev Lett* **2011,** *107* (12), 127402.
63. Yin, W.-J.; Yang, J.-H.; Kang, J.; Yan, Y.; Wei, S.-H., Halide perovskite materials for solar cells: a theoretical review. *Journal of Materials Chemistry A* **2015,** *3* (17), 8926-8942.